\documentclass[prd,nofootinbib,showpacs,12pt]{revtex4}

\usepackage[dvips]{color,graphicx}
\usepackage{subfigure}

\def\t{{\rm tot}}
\def\beq{\begin{equation}}
\def\eeq{\end{equation}}
\def\ber{\begin{eqnarray}}
\def\eer{\end{eqnarray}}

\def\prd{{Phys.\@ Rev.\@ D\ }}
\def\pd{{Phys.\@ Rev.\@ D\ }}

\def\plb {{Phys.\@ Lett.\@ B\ }}
\def \jetpl {JETP Lett.\ }

\def\ie {{i.e.}}

\begin{document}

\title{Braneworld dynamics in Einstein--Gauss--Bonnet gravity}

\author{Hideki Maeda}\email{hideki@cecs.cl}
\affiliation{Centro de Estudios Cient\'{\i}ficos (CECS), Arturo Prat 514, Valdivia, Chile}
\affiliation{Department of Physics, International Christian University, 3-10-2 Osawa, Mitaka-shi,
Tokyo 181-8585, Japan}
\author{Varun Sahni}\email{varun@iucaa.ernet.in}
\affiliation{Inter-University Centre for Astronomy \& Astrophysics, Post Bag 4,
Pune~411~007, India}
\author{Yuri Shtanov}\email{shtanov@bitp.kiev.ua}
\affiliation{Bogolyubov Institute for Theoretical Physics, Kiev 03680, Ukraine}

\date{\today}

\begin{abstract}
We discuss the cosmological evolution of a braneworld in five dimensional
Gauss--Bonnet gravity. Our discussion allows the fifth (bulk) dimension to be
space-like as well as time-like. The resulting equations of motion have the
form of a cubic equation in the $\left(H^2,\,  (\rho+\sigma)^2\right)$ plane,
where $\sigma$ is the brane tension and $\rho$ is the matter density. This
allows us to conduct a comprehensive pictorial analysis of cosmological
evolution for the Gauss--Bonnet brane. The many interesting properties of this
braneworld include the possibility of accelerated expansion at late times. For
a finite region in parameter space the accelerated expansion can be {\em
phantom-like\/} so that $w < -1$. At late times, this branch approaches
de~Sitter space ($w=-1$) and avoids the {\em big-rip\/} singularities usually
present in phantom models. For a time-like extra dimension the Gauss--Bonnet
brane can bounce and avoid the initial singularity.
\end{abstract}

\pacs{04.50.+h, 98.80.Cq}

\maketitle

\section{introduction}
Braneworld models of the universe --- in which the observable universe is a
four dimensional timelike hypersurface (brane) embedded in a higher dimensional
(bulk) space-time --- have attracted much recent attention. This is partly due
to the fact that Superstring/M-theory seems to require the existence of extra
dimensions and the braneworld approach may be one way of reconciling our 3+1
dimensional universe with these higher dimensional theories
\cite{Lukas,large,Randall}.

Another reason for the current popularity of the braneworld construct is due to
the fact that brane cosmology is usually accompanied by new features and is
therefore, in principle, falsifiable \cite{maartens04,sahni05}. The simplest
Randall--Sundrum (RS) braneworld, for instance, gives rise to an evolutionary
equation for the brane which differs from standard general relativity at {\em
early times\/} \cite{Randall}. This leads to several interesting consequences.
For instance, the very early universe expands as $H \propto \rho$, instead of
the more familar $H \propto \sqrt{\rho}$ in standard cosmology. The changed
expansion rate causes a scalar field to experience greater damping, which, in
turn, allows Inflation to occur for a broader class of initial conditions and
potentials \cite{copeland01}. If the fifth dimension is timelike then the
universe generically {\em bounces\/} and avoids the initial big bang
singularity which plagues standard cosmology \cite{ss03b}. (The behaviour of
anisotropies in the RS scenario can also be very different from that in
standard general relativity \cite{mss01}.)

A complementary approach to braneworld cosmology pioneered by the DGP model
\cite{DGP}, allows the universe to accelerate at {\em late times\/} thus
providing a geometrical answer to the riddle posed by dark energy. Models which
unify the RS and DGP approaches also lead to several new features
\cite{ss03a,sahni05}. For instance (i) the phenomenon of dark energy can be
transient so that the universe accelerates for a while before settling back
into matter dominated expansion, (ii) the effective equation of state of dark
energy can be {\em phantom-like\/} ($w_{\rm eff} \leq -1$), (iii) new
cosmological singularities can arise in such models \cite{ss02}. Such
alternative cosmological models provide reasonable fits to the current
cosmological data \cite{alam06}.

In this paper we address the issue of cosmological evolution on a brane in  a
theory of gravity whose action includes, in addition to the familiar Einstein
term, a Gauss--Bonnet contribution. Gauss--Bonnet terms arise naturally in
superstring theories \cite{Gross} and their cosmological effects have been
discussed in several papers \cite{GB,gw2003,davis2003}.  The present paper
deals with this issue in greater generality, we examine both cases: when the
bulk dimension is spacelike as well as timelike. We also a develop a new
pictorial method of analysis which provides qualitative insights into the
evolution of the universe in this potentially important new model of gravity.

\section{Basic equations}
We begin with the following $n$-dimensional ($n \geq 5$) action:
\begin{equation}
\label{action} S=\int
d^nx\sqrt{-g}\biggl[\frac{1}{2\kappa_n^2}(R-2\Lambda+\alpha{L}_{GB}) \biggr] \,
,
\end{equation}
where $R$ is the $n$-dimensional Ricci scalar, $\Lambda$ is the $n$-dimensional
cosmological constant, and $\kappa_n:=\sqrt{8\pi G_n}$, where $G_n$ is the
$n$-dimensional gravitational constant. The Gauss--Bonnet term ${L}_{GB}$ is a
combination of the Ricci scalar, the Ricci tensor $R_{\mu\nu}$ and the Riemann
tensor $R^\mu{}_{\nu\rho\sigma}$:
\begin{equation}
{L}_{GB}:=R^2-4R_{\mu\nu}R^{\mu\nu}+R_{\mu\nu\rho\sigma}R^{\mu\nu\rho\sigma}\,
.
\end{equation}
The constant $\alpha$ in (\ref{action}) is the coupling constant of the
Gauss--Bonnet term and for $\alpha\to 0$ our model reduces to the familiar
Randall--Sundrum model \cite{Randall}. The action (\ref{action}) can be
obtained in the low-energy limit of heterotic superstring theory~\cite{Gross},
in which case $\alpha$ can be regarded as the inverse string tension and is
positive-definite. We, therefore, assume $\alpha > 0$ throughout this paper.
(We shall explicitely be assuming $n \geq 5$ since for $n \le 4$ the
Gauss--Bonnet term is a topological invariant and does not contribute to the
field equations.)

The gravitational equations which result from the action (\ref{action}) are
\begin{equation}
{G}^\mu{}_\nu +\alpha {H}^\mu{}_\nu +\Lambda \delta^\mu{}_\nu= 0 \, ,
\label{beq}
\end{equation}
where
\begin{eqnarray}
{G}_{\mu\nu}&:=&R_{\mu\nu}-{1\over 2}g_{\mu\nu}R\, ,\\
{H}_{\mu\nu}&:=&2\Bigl[RR_{\mu\nu}-2R_{\mu\alpha} R^\alpha{}_\nu -
2R^{\alpha\beta} R_{\mu\alpha\nu\beta} + R_\mu{}^{\alpha\beta\gamma}
R_{\nu\alpha\beta\gamma} \Bigr] - {1\over 2} g_{\mu\nu} {L}_{GB}\, .
\end{eqnarray}

\subsection{Bulk solution}
\label{bulk}
The $n$-dimensional vacuum solution can be obtained as a product manifold
$M^n \approx M^2\times K^{n-2}$ with the line element
\begin{equation}
ds_n^2=-h(r)dt^2+\varepsilon\frac{dr^2}{h(r)}+r^2\gamma_{ij}dx^idx^j \, ,
\label{eq:product_manifold}
\end{equation}
where $K^{n-2}$ is an $(n-2)$-dimensional space of constant curvature with unit
metric $\gamma_{ij}$. In the equations which follow, $k$ denotes the curvature
of $K^{n-2}$ and takes the values $1$ (positive curvature), $0$ (zero
curvature), and $-1$ (negative curvature). The value of the constant
$\varepsilon$ determines whether the (bulk) fifth dimension is spacelike
($\varepsilon = 1$) or timelike ($\varepsilon = -1$). In the former case, $M^2$
is a Lorenzian manifold, whereas in the latter case, it is a Euclidean
manifold.

The basic equations of the theory under consideration are
\begin{eqnarray}
0&=&r^2\biggl[2\alpha(n-3)(n-4)h-\varepsilon \biggl\{r^2+2\alpha k(n-3)(n-4)\biggl\}\biggl]\frac{dh^2}{dr^2} \nonumber \\
&&+2(n-3)r\biggl[2\alpha(n-4)(n-5)h-\varepsilon \biggl\{r^2+2\alpha k(n-4)(n-5)\biggl\}\biggl]\frac{dh}{dr} \nonumber \\
&&+2\alpha r^2(n-3)(n-4)\biggl(\frac{dh}{dr}\biggl)^2+\alpha(n-3)(n-4)(n-5)(n-6)h^2 \nonumber \\
&&-\varepsilon (n-3)(n-4)h\biggl[r^2+2\alpha k (n-5)(n-6)\biggl] \nonumber \\
&&-2\Lambda r^4+k(n-3)(n-4)r^2+\alpha k^2(n-3)(n-4)(n-5)(n-6)\, ,\\ \nonumber \\
0&=&(n-2)r\biggl[2\alpha(n-3)(n-4)h-\varepsilon \biggl\{r^2+2\alpha k(n-3)(n-4)\biggl\}\biggl]\frac{dh}{dr} \nonumber \\
&&+\alpha(n-2)(n-3)(n-4)(n-5)h^2 \nonumber \\
&&-\varepsilon (n-2)(n-3)h\biggl[r^2+2\alpha k (n-4)(n-5)\biggl] \nonumber \\
&&-2\Lambda r^4+k(n-2)(n-3)r^2+\alpha k^2(n-2)(n-3)(n-4)(n-5)\, ,
\end{eqnarray}
where the former is the $(i,i)$ component of Eq.~(\ref{beq}), while the latter
is the $(t,t)$ or $(r,r)$ component acting as a constraint. The general
solution of these equations is
\begin{equation}
h(r)= \varepsilon k+\frac{r^2}{2(n-3)(n-4)\alpha}
\left(\varepsilon\mp\sqrt{1+\frac{\alpha\mu}{r^{n-1}}+\frac{8(n-3)(n-4)}{(n-1)(n-2)}
\alpha \Lambda}\right) \, , \label{horizon}
\end{equation}
where $\mu$ is a constant.
Our solution for $h(r)$ has two branches,
which correspond to the two signs in front of the square root in Eq.~(\ref{horizon}).
We call the family with the minus (plus) sign the minus-branch (plus-branch) solution.

\begin{itemize}

\item For $\varepsilon=1$, the minus-branch solution has the general relativistic
limit as $\alpha \to 0$, while there is no general relativistic limit for the
plus-branch solution. (The global structures of this solution were presented
in~\cite{tm2005}.)

\item For $\varepsilon=-1$, the plus-branch solution has the general relativistic limit
as $\alpha \to 0$, while the minus-branch solution does not.

\end{itemize}
Hereafter, we shall be considering a five-dimensional bulk spacetime, for which
the metric (\ref{eq:product_manifold}) reduces to
\begin{eqnarray}
\label{h-eq} ds_5^2&=&g_{\mu\nu}dx^\mu dx^\nu
= -h(r)dt^2+\varepsilon\frac{dr^2}{h(r)}+r^2 \left[
d\chi^2+f_k(\chi)^2(d\theta^2+\sin^2\theta d\phi^2) \right]\, ,\\
h(r) &=& \varepsilon k+\frac{r^2}{4\alpha}\Biggl(\varepsilon\mp
\sqrt{1+\frac{\alpha \mu}{r^{4}}+\frac43\alpha\Lambda}\Biggr) \,
,\label{horizon-2}
\end{eqnarray}
where $f_0(\chi)=\chi$, $f_1(\chi)=\sin\chi$,  $f_{-1}(\chi)=\sinh\chi$, and
$\varepsilon=\pm1$.

In this spacetime, there are two classes of singularities when $\mu \ne 0$. One
is the central singularity at $r=0$ and the other is the {\it branch
singularity\/} at $r=r_{\rm b} := [-\alpha\mu/(1+4\alpha\Lambda/3)]^{1/4}> 0$,
when the term inside the square-root in Eq.~(\ref{horizon-2}) vanishes. The
branch singularity exists if $\mu$ is negative, or if $1+4\alpha\Lambda/3<0$
for positive $\mu$.

\subsection{Friedmann equation on the brane}
The position of the three-brane is described by the functions $r=a(\tau)$ and
$t=T(\tau)$ parametrized by the proper time $\tau$ on the brane. The tangent
vector to the brane is written as
\begin{equation}
u^\mu\frac{\partial}{\partial x^\mu}={\dot T} \frac{\partial}{\partial t}+{\dot
a}\frac{\partial}{\partial r}\, ,
\end{equation}
where a dot denotes the differentiation with respect to $\tau$.
The normalization condition $u_\mu u^\mu=-1$ leads to
\begin{equation}
1=h(a){\dot T}^2-\varepsilon\frac{{\dot a}^2}{h(a)}\, , \label{norm}
\end{equation}
and the induced metric of the three-brane ${\bar g_{ab}}$ is given by
\begin{equation}
\label{1stout} ds_4^2={\bar g_{ab}}dy^ady^b= -d\tau^2+a(\tau)^2 \left[
d\chi^2+f_k(\chi)^2(d\theta^2+\sin^2\theta d\phi^2) \right]\, .
\end{equation}
The unit normal 1-form to the three-brane $n_\mu$ is given by
\begin{equation}
n_\mu dx^\mu ={\dot a}dt-{\dot T}dr\, ,
\end{equation}
where $n_\mu u^\mu=0$ and $n_\mu n^\mu=1/\varepsilon$ are satisfied.

The extrinsic curvature of the three-brane is obtained from $K_{ab}:=
n_{\mu;\nu}e^\mu_a e^\nu_b$, where $e^\mu_a := \partial x^\mu/\partial y^a$. We
have
\begin{equation}
e^0_ady^a ={\dot t}d\tau \, , \qquad e^1_ady^a = {\dot a}d\tau \, , \qquad
e^i_ady^a = \delta^i_j dy^j \, ,
\end{equation}
and
\begin{equation}
K_{ab}=-n_{\mu}e_{a,b}^\mu-\Gamma^\kappa_{\mu\nu}n_\kappa e_a^\mu e_b^\nu \, .
\end{equation}
Then, we obtain the non-zero component of $K^a{}_b$ as
\begin{equation}
K^\tau{}_\tau = -\frac{1}{h{\dot T}}\left({\ddot
a}+\frac{h'}{2\varepsilon}\right)\, , \qquad K^i{}_j = -\frac{h{\dot
T}}{\varepsilon a}\delta^i{}_j \, ,
\end{equation}
where a prime denotes differentiation with respect to $a$.

The junction condition at the brane is given by \cite{gw2003,davis2003,gggw2008}
\begin{equation} \label{j-condition}
[K^a{}_b]_\pm - \delta^a{}_b [K]_\pm + 2\alpha \left( 3\varepsilon[J^a{}_b]_\pm
-\varepsilon\delta^a{}_b [J]_\pm - 2 P^a{}_{dbf}[K^{df}]_\pm \right) = -
\varepsilon\kappa_5^2 \tau^a{}_b \, ,
\end{equation}
where
\begin{eqnarray}
J_{ab} &:=&{1\over 3} \left(2KK_{ad}K^{d}{}_{b}+K_{df}K^{df}K_{ab}
-2K_{ad}K^{df}K_{fb}-K^2 K_{ab}\right) \, , \\
P_{adbf}&:=&R_{adbf}+2h_{a[f}R_{b]d}+2h_{d[b}R_{f]a} +Rh_{a[b}h_{f]d} \, .
\end{eqnarray}
The energy-momentum tensor $\tau^a{}_b$ on the brane is given by
\begin{equation}
\tau^a{}_b = \mbox{diag} (-\rho,p,p,p) + \mbox{diag}
(-\sigma,-\sigma,-\sigma,-\sigma,) \, ,
\end{equation}
where $\rho$ and $p$ are the energy density and pressure of a perfect fluid on
the three-brane, and the constant $\sigma$ is the brane tension. We have
introduced the notation
\begin{equation}
[X]_\pm:= X^+-X^- \, ,
\end{equation}
where $X^\pm$ is the quantity $X$ evaluated either on the $+$ or $-$ side of
the brane, and $P_{adbf}$ is the divergence-free part of the Riemann tensor,
i.e.,
\begin{equation}
D_a P^{a}{}_{dbf}=0 \, ,
\end{equation}
where $D_a$ is the covariant derivative on the brane.

From the $(\tau,\tau)$ and $(i,i)$ components of Eq.~(\ref{j-condition}) and
Eq.~(\ref{norm}), we obtain
\begin{equation}
\frac{\kappa^4_5}{36}(\rho+\sigma)^2 = \left(\frac{h(a)}{a^2}+\varepsilon
H^2\right)\left[1+\frac{4\alpha}{3}
\left(\frac{3k-\varepsilon h(a)}{a^2}+2H^2\right)\right]^2\, , \label{GB-F}
\end{equation}
where $H := {\dot a}/a$. Here, we have assumed $Z_2$-symmetry of reflection with respect
to the brane. This generalized Friedmann equation reduces to that obtained by Davis
\cite{davis2003} for $\varepsilon=1$.

Differentiating Eq.~(\ref{GB-F}) with respect to $\tau$ and using
Eq.~(\ref{norm}) and the $(\tau,\tau)$ and $(i,i)$ components of
Eq.~(\ref{j-condition}), we obtain
\begin{equation}
{\dot \rho}=-3H(p+\rho) \, , \label{em-cons}
\end{equation}
which is the energy-conservation equation on the three-brane.
Let us assume that the perfect fluid on the three-brane obeys
\begin{equation}
p=(\gamma-1)\rho \, , \label{eos}
\end{equation}
where we assume that the equation of state of matter on the brane lies within
the Zeldovich interval $0<\gamma \le 2$ due to the dominant energy condition
(equivalently, $-1 < w \leq 1$, where $w := p/\rho = 1+\gamma$ is the equation
of state). From Eq.~(\ref{em-cons}), we then obtain
\begin{equation}
\rho=\frac{\rho_0}{a^{3\gamma}}\, , \label{energy}
\end{equation}
where $\rho_0$ is a positive constant, so that $\rho$ is a monotonically
decreasing function of $a$ for $\gamma>0$.

\subsection{The Randall--Sundrum limit}

In this paper, we shall consider only those solutions of (\ref{horizon-2}) and
(\ref{GB-F}) which possess the general-relativistic limit since other solutions
may describe physically inadmissible evolution of our brane. The minus-
and plus-branch solutions in (\ref{horizon-2}) have the general relativistic
limits for $\varepsilon=1$ and $-1$ in (\ref{GB-F}), respectively.

As mentioned earlier, the action (\ref{action}) contains the Randall--Sundrum
model as a subclass. Setting $\alpha \to 0$ in Eq.~(\ref{GB-F}), one gets the
generalised Randall--Sundrum (RS) model
\begin{eqnarray}
\frac{\kappa^4_5}{36}(\rho+\sigma)^2 =
\frac{\varepsilon}{a^2}\left(k-\frac{\mu}{8 a^2}-\frac16\Lambda
a^2\right)+\varepsilon H^2\, . \label{GR-F}
\end{eqnarray}
From this equation and from Eq.~(\ref{energy}), we obtain
\begin{eqnarray}
H^2=\frac{\kappa^4_5}{36\varepsilon}\left(\frac{\rho_0}{a^{3\gamma}} +
\sigma\right)^2-\frac{k}{a^2}+\frac{\mu}{8 a^4}+\frac16\Lambda\, .
\label{bounce}
\end{eqnarray}
The Randall--Sundrum model corresponds to $\varepsilon = 1$, while the dual
model with $\varepsilon=-1$ was discussed in \cite{ss03b}. Permitted values of
the expansion factor must clearly satisfy $H^2 \geq 0$. An interesting
consequence of (\ref{bounce}) is the possibility of singularity-free solutions
when $\varepsilon=-1$ \cite{ss03b}.

\section{Pictorial analysis of cosmological evolution}

We saw in the previous section that the evolution equation for the
Gauss--Bonnet brane can be quite complicated and, therefore, difficult to
analyze analytically. In this section, we present a general method of analysis
which allows one to study pictorially the behaviour arising from the generic
cosmological equation (\ref{GB-F}).

We notice that equation (\ref{GB-F}), describing the cosmological evolution of
the Gauss--Bonnet brane, always has the form of a {\em cubic curve} in the
$\left(H^2,\, \rho_\t^2\right)$ plane:
\begin{equation} \label{cubic_curve}
C \rho_\t^2 = \left(A \pm H^2\right) \left( B + H^2 \right)^2 \, ,
\end{equation}
where $\rho_\t :=\rho + \sigma$, $A$ and $B$ are functions of $a$, $C$ is a
positive constant and the $\pm$ sign corresponds to $\varepsilon = \pm 1$. The
value of cosmological constant $\sigma$ can be positive, negative or zero.  The
right-hand side of equation (\ref{cubic_curve}) has exactly three real zeros in
$H^2$, two of which coincide, namely, $\left(H^2\right)_1 = \mp A$, and
$\left(H^2\right)_{2,3} = - B$. Only part of this cubic curve lies in the
physical domain $H^2 \ge 0$, $\rho_\t^2 \ge 0$, and it is in this domain that
the evolution of the brane takes place. Consequently, the evolution of our
brane-universe can be pictured as a point moving along this cubic curve in the
physical domain $\rho_\t^2 \ge 0$, $H^2 \ge 0$.

This pictorial representation is very useful in appreciating the full gamut of
possibilities for cosmic evolution of this brane. For comparison, it is helpful
to note that cosmological evolution in general relativity (GR) is described by
\beq
H^2 = \rho_\t - \frac{k}{a^2}\, , \label{eq:quadratic}
\eeq
where we have set the proportionality term $8\pi G/3$ to unity. Equation
(\ref{eq:quadratic}) describes a {\em quadratic curve\/} in the $\left(H^2,\,
\rho_\t^2\right)$ plane. Another example is the Randall--Sundrum brane, which,
for every value of $a$, is described by a {\em straight line\/} in the
$\left(H^2,\, \rho_\t^2\right)$ plane:
\beq \label{eq:RS_bounce}
H^2 = \frac{\rho_\t^2}{\varepsilon} + \frac{\Lambda}{6} - \frac{k}{a^2} +
\frac{\mu}{8a^4}\, ,
\eeq
where $\varepsilon = \pm 1$, and we have set the term $\kappa_5^4/36$ in
(\ref{bounce}) to unity.

\begin{figure}[htbp]
\begin{center}
\includegraphics[width=0.5\linewidth]{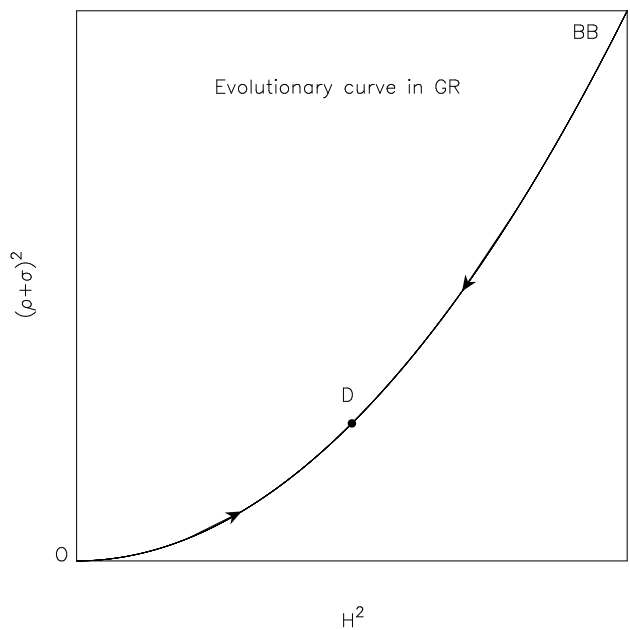}
\caption{\label{fig:GR} The evolution of a spatially flat FRW universe in GR
proceeds along this curve. The downward arrow indicates {\em expansion\/} while
the upward arrow indicates {\em contraction\/}. The latter is only possible if
$\sigma < 0$. For $\sigma > 0$ the expansion of the universe terminates at the
point $D$ at which $\rho = 0$. At this point, the universe expands
exponentially. For $\sigma = 0$, the origin ($H=0$, $\rho=0$) marks the end
point of evolution. The scale of the $x$ and $y$-axis is arbitrary. }
\end{center}
\end{figure}

Before commencing our discussion on the subtleties of cosmological evolution on
the Gauss--Bonnet brane, it will be helpful to first consider the different
evolutionary possibilities in a spatially flat universe described by the more
familiar general-relativistic equation (\ref{eq:quadratic}) with $k = 0$, where
$\sigma$ acts as a cosmological constant. In this case the expansion of the
universe can proceed in three distinct ways, corresponding to the cases $\sigma
> 0$, $\sigma < 0$ and $\sigma = 0$. All three possibilities correspond to
motion along the quadratic curve in Fig.~\ref{fig:GR}.

Notice that expansion along the {\em entire curve\/} from the top (early times)
to the origin (late times) takes place only if $\sigma \leq 0$. In the absence
of a cosmological constant ($\sigma = 0$) the origin $\left(H^2,\,
\rho_\t^2\right) = (0,0)$ is reached at $\tau \to \infty$. In contrast, for
$\sigma < 0$, the origin is reached in a finite interval of time when the
matter density has dropped to $\rho = |\sigma|$. At this point $H = 0$, in
other words expansion ceases and the universe begins to contract. Evolution
thereafter proceeds upward along the same curve --- in {\em reverse\/} fashion.

Finally, if $\sigma > 0$, evolution does not proceed all the way to the origin
but terminates at some point $D$ along the curve. At this point, $\rho \to 0$
so that $\rho_\t = \sigma$ and $H^2 = \sigma$. The end point $D$ of evolution
corresponds to the universe's asymptotic approach towards de~Sitter space.
(This, for instance, would be the case for a spatially flat $\Lambda$CDM
universe which accelerates at late times.)

\subsection{Spacelike extra dimension ($\varepsilon=1$)}

Let us now discuss the evolution on the Gauss--Bonnet brane in greater detail.
For a spacelike extra dimension, the cosmological equation (\ref{GB-F}) has the
form
\begin{equation} \label{plus}
C \rho_\t^2 = \left(A + H^2\right) \left( B + H^2 \right)^2 \, ,
\end{equation}
where
\begin{equation}
C := {\kappa_5^4 \over 36} \left( {3 \over 8 \alpha} \right)^2 > 0 \, ,
\end{equation}
and
\begin{equation} \label{ab-plus}
A := {1 \over 4 \alpha} \left(1 \mp \sqrt{1 + {\alpha \mu \over a^4} + \frac43
\alpha \Lambda} \right) \, , \quad B := {1 \over 8 \alpha} \left(2 \pm \sqrt{1
+ {\alpha \mu \over a^4} + \frac43 \alpha \Lambda} \right) = {3 \over 8 \alpha}
- \frac{A}{2} \, ,
\end{equation}
in general, are functions of the scale factor $a$.

As mentioned earlier, equation (\ref{plus}) has the form of a cubic curve in the
$\left(H^2,\,  \rho_\t^2\right)$ plane. The two signs in (\ref{ab-plus})
correspond to the two different ways of embedding the brane in the bulk space.
In this paper we only consider the upper sign, which has the GR limit.

As discussed in the previous section, the evolution of the braneworld is
described by a point moving along the cubic curve in the $\left(H^2,\,
\rho_\t^2\right)$ plane, in the physical domain $H^2 \ge 0$, $\rho_\t^2 \ge 0$,
with the parameters of the cubic curve simultaneously changing with time due to
the dependence of $A$ and $B$ on the scale factor (see below).  The evolution
can proceed in three distinct ways which are summarized below. All three
cases correspond to $B > 0$ in equation (\ref{ab-plus}), and the first two also
have $A < 0$.

\begin{enumerate}

\item
The behaviour of the universe is shown in the left panel of
Fig.~\ref{fig-space-pp}.  The point $P$ corresponds to $H^2 = - A$.  During the
course of expansion, the motion along the curve is initially downwards from the
initial Big Bang (BB) singularity towards $P$. However, for $P$ to be reachable
in a finite time interval the brane tension $\sigma$ must be negative since
only then is $(\rho + \sigma)^2 = 0$ permitted. The point $P$ marks a turning
point for the evolution along the curve: after this point, the energy density
of the universe keeps decreasing while the quantity $(\rho + \sigma)^2$ is
increasing.  In the case $\mu = 0$, we also have ${\dot H} = 0$ at the point
$P$. In this case, the Hubble parameter passes through an inflection point at
$P$. Since
\beq
\frac{\ddot a}{a} = {\dot H} + H^2 \, ,
\eeq
it follows that ${\ddot a} = aH^2 > 0$ when $\rho = |\sigma|$. In other words,
${\dot H} > 0$ for some length of time during the upward motion along the curve
{\em away from\/} $P$. Thus the universe {\em accelerates at late times\/}.
Note that ${\dot H} = -4\pi G\rho$ in $\Lambda$CDM and ${\dot H} > 0$ is
usually associated with a {\em phantom\/} equation of state $w < -1$ in
dark-energy models.  (This qualitative behaviour will take place also for
sufficiently small values of $\alpha\mu/a^4$ reached in the course of expansion
in the neighbourhood of the point $P$, which will make $A$ almost constant in
time.)

The growth in $H$, however, cannot continue indefinitely since $\rho \to 0$
eventually, and $(\rho + \sigma)^2 \to \sigma^2$ (corresponding to the point
$D$), which implies $H^2 \to \mbox{const}$. This means that the universe
approaches a de~Sitter-like attractor ($D$) at very late times. We therefore
conclude that our Gauss--Bonnet brane can display phantom-like features and
super-accelerate at late times, before approaching $w \to -1$ in the distant
future. Note that the big-rip future singularity (at which $H \to \infty$) is
absent in this case, which is one of the appealing features of this scenario.
(Other braneworld models with this property have been discussed in
\cite{ss03a}.)

\item
For $\sigma > 0$, the point $P$ can never be reached, and expansion proceeds
along $BB \to D$, culminating in de~Sitter-like asymptotic expansion at $D$.

\begin{figure}[htbp]
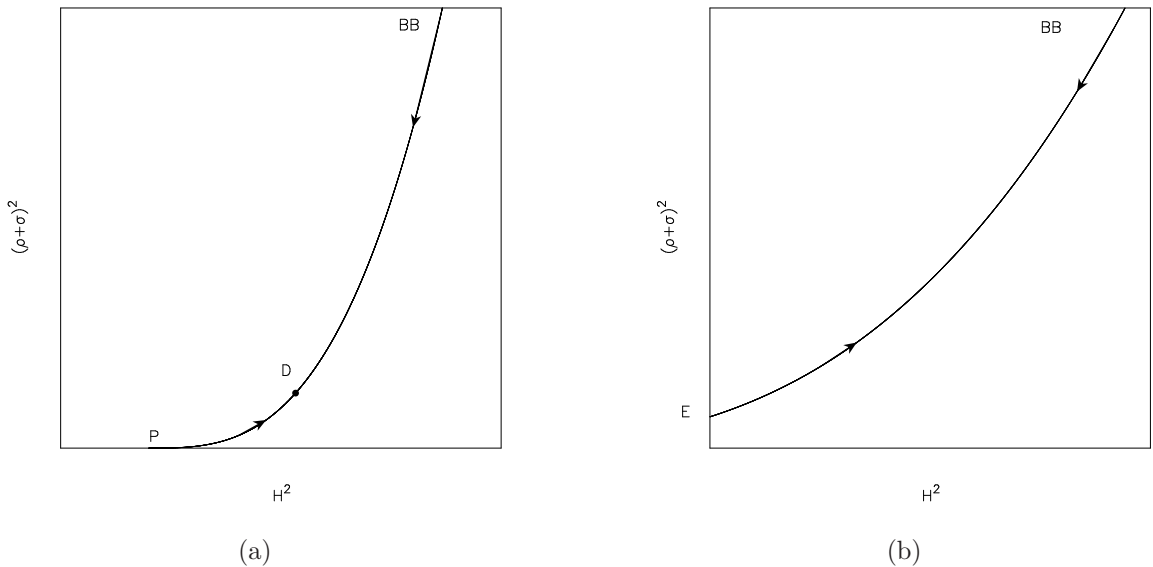

\begin{center}
\subfigure[]{\includegraphics[width=0.4\linewidth]{space-mp.ps}}~~~~~~~~~~~~~~~
\subfigure[]{\includegraphics[width=0.4\linewidth]{space-pp.ps}}
\caption{\label{fig-space-pp}
Spacelike extra dimension: $A < 0$, $B > 0$ (left) and $A > 0$, $B > 0$ (right)
in (\ref{plus}).  The point $P$ is the turning point, and the point $E$ is the
point of recollapse.
}
\end{center}
\end{figure}

\item The case with $B > 0$ and $A > 0$ is shown in the right panel of
Fig.~\ref{fig-space-pp}.  The point $E$ is the point of recollapse. At this
point the (spatially flat) universe ceases to expand and begins to contract.
The point $E$ is reachable either if the brane tension is negative, or if it is
positive with the value of $\sigma^2$ lying below the point $E$.
\end{enumerate}

One should note that, theoretically, the scale-factor dependent parameter
$A(a)$ can change sign during the course of evolution, so that the curve along
which the evolution takes place can continuously evolve from that in the right
panel of Fig.~\ref{fig-space-pp} to that in its left panel, and vice versa.
This introduces an obvious modification to the description of the evolution,
which does not change in any significant way.

The complete set of figures showing the $(H^2,\, \rho_{\rm tot}^2)$ plane are
shown in Fig.~\ref{fig:spacelike-all} of the Appendix.

\subsection{Timelike extra dimension ($\varepsilon=-1$)}

Also in this case, there is only one branch of the generic cosmological
equation (\ref{GB-F}) having the GR limit which has the form
(\ref{cubic_curve}), namely,
\begin{equation} \label{minus}
C \rho_\t^2 = \left(A - H^2 \right) \left( B + H^2 \right)^2 \, ,
\end{equation}
where
\begin{equation}
C:={\kappa_5^4 \over 36} \left( {3 \over 8 \alpha} \right)^2 > 0 \, ,
\end{equation}
and
\begin{equation} \label{ab-minus}
A:={1 \over 4 \alpha} \left(\sqrt{1 + {\alpha \mu \over a^4} + \frac43 \alpha
\Lambda} - 1 \right) \, , \quad B:={1 \over 4 \alpha} \left(1 + \frac12
\sqrt{1 + {\alpha \mu \over a^4} + \frac43 \alpha \Lambda} \right) = {3 \over
8 \alpha} + \frac{A}{2} \, .
\end{equation}

\begin{figure}[htbp]
\begin{center}
\subfigure[]{\includegraphics[width=0.45\linewidth]{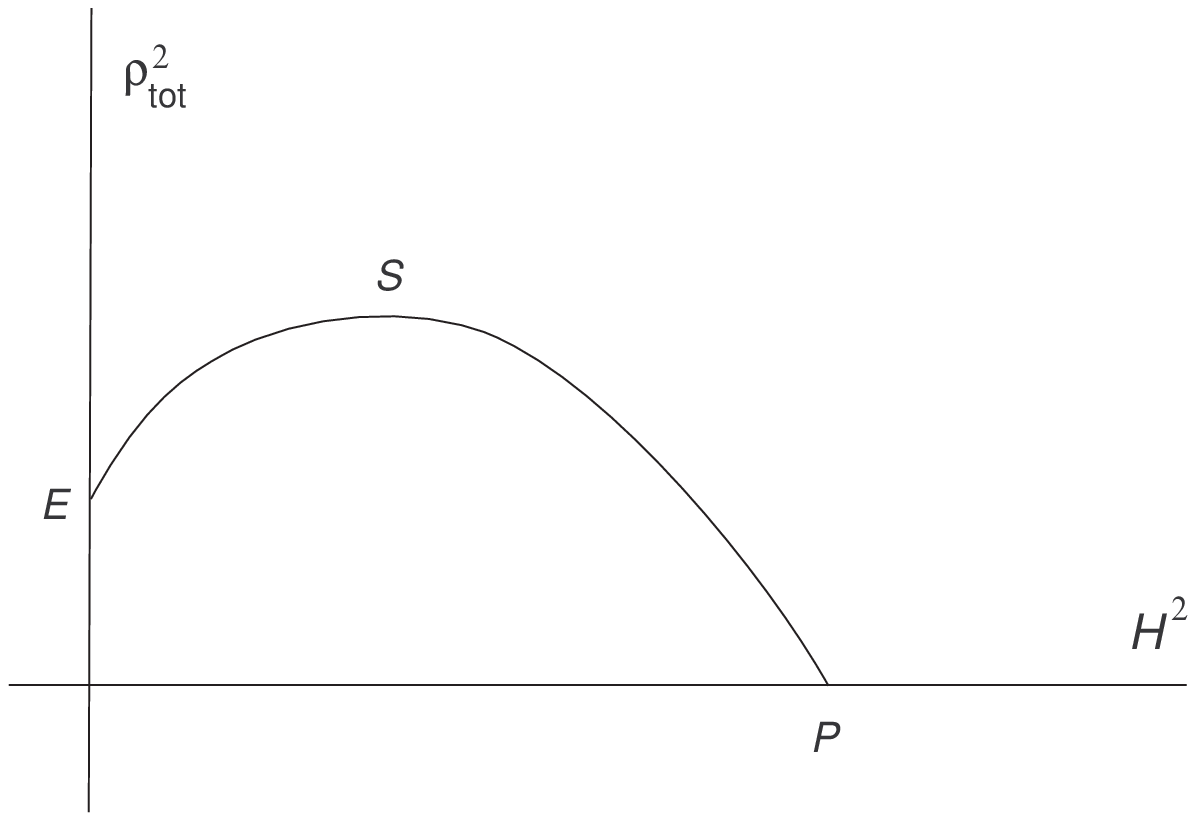}}~~~~~~~~~~~~~~~
\subfigure[]{\includegraphics[width=0.45\linewidth]{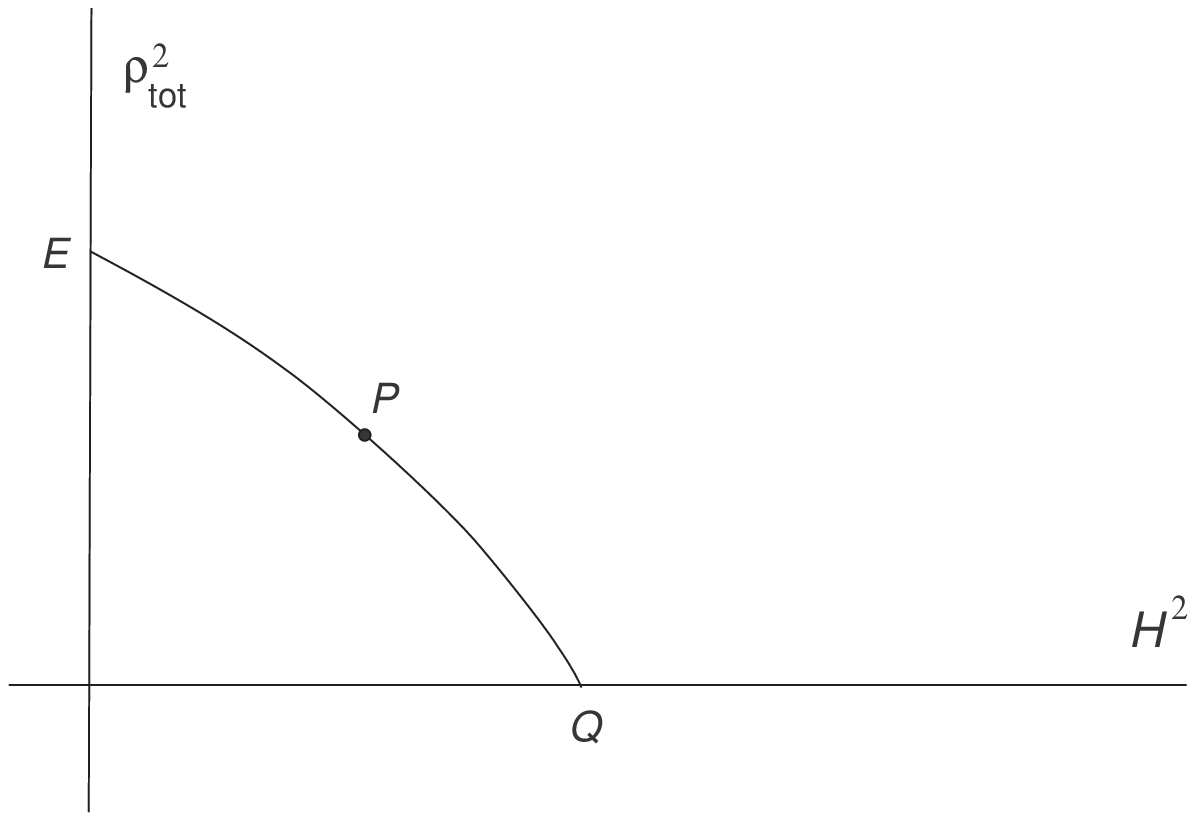}}
\caption{\label{fig-time1}
Gauss--Bonnet brane with timelike extra dimension: \\
(a)~$0 < B < 2A$.  The points $E$ (bounce or recollapse) and $P$ are the turning points
of the evolution, while the point $S$ corresponds to a sudden or `quiescent' singularity.
\\ (b)~$B > 2A$. The bouncing scenario (which requires $\sigma < 0$) describes a brane contracting
from a de~Sitter-like initial stage at $P$ to $Q$ at which
$\rho = \vert\sigma\vert$ and $\rho_{\rm tot} := \rho + \sigma = 0$.
Further contraction takes the universe from $Q$ to $E$, and along this segment both
$\rho$ as well as $\rho+\sigma$ increase. At $E$, the density of the universe has reached
its maximum value while the Hubble parameter has declined to zero. The universe therefore
bounces at $E$, then re-expands and evolves in reverse fashion along $E\to Q\to P$. Note
that $P$ marks the beginning and end point of evolution.}
\end{center}
\end{figure}

Clearly, the theory makes sense only for $A > 0$ (for which the branch singularity does
not appear), hence, also $B > 0$. A typical graph illustrating the case $0 < B < 2A$
(equivalently $1 /4 \alpha < A$) is shown in Fig.~\ref{fig-time1}(a). The point $P$
corresponds to $H^2 = A$. The graph corresponding to $B > 2A$, which is equivalent to
\begin{equation}
A < {1 \over 4 \alpha} \, , \quad \mbox{or} \quad \sqrt{1 + {\alpha \mu \over
a^4} + \frac43 \alpha \Lambda} < 2 \, ,
\end{equation}
is shown in Fig.~\ref{fig-time1}(b).

The end points $E$ and $P$ in Fig.~\ref{fig-time1}(a) are the reverse points of
evolution. The point $S$ is the position of a sudden `quiescent' singularity of the type
described in \cite{ss02,barrow}. Indeed, the evolution of the universe cannot be
continued beyond this point because the quantity $\rho_\t^2$ should change in the same
direction (decrease), which is physically impossible. Note that the value of $H$ is
finite and nonzero at this point, while $\dot H$ is divergent. (This can easily be seen
by writing $d(H^2)/d\tau = d(H^2)/d\rho_{\rm tot}\cdot d\rho_{\rm tot}/d\tau$ where
$d\rho_{\rm tot}/\tau = -3H\rho$ and noting that $d(H^2)/d\rho_{\rm tot} \to \infty$ at
$S$.) The Kretschmann invariant on the brane $K:=R_{abcd}R^{abcd}$ is given by
\begin{equation}
K=12[H^4+({\dot H}+H^2)^2]~,
\end{equation}
and diverges as the {\em quiescent\/} singularity is approached.

\begin{figure}[htbp]
\begin{center}
\includegraphics[width=0.4\linewidth]{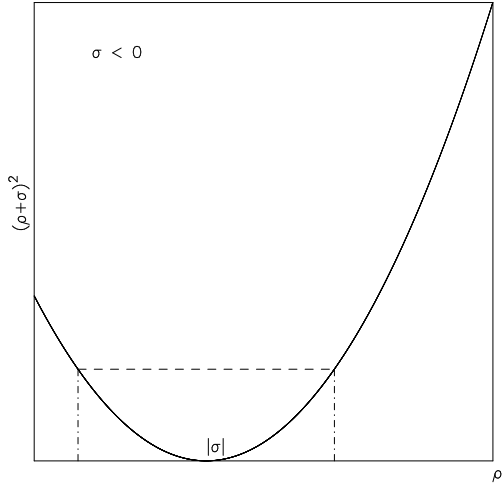}
\caption{\label{fig-map}
For negative values of the brane tension ($\sigma < 0$) two
values of the matter density $\rho$ map onto a single value of $\rho_{\rm tot}
= \rho+\sigma$, as illustrated in this figure.
}
\end{center}
\end{figure}

Consider now in more detail the evolution of the Gauss--Bonnet brane suggested by
Fig.~\ref{fig-time1}(a). The Big Bang singularity which featured prominently in
Fig.~\ref{fig-space-pp} has effectively been replaced by the sudden singularity $S$. The
following four possibilities for evolution immediately suggest themselves:

\begin{enumerate}

\item Expansion commences at $S$ and proceeds to $E$, which marks a turning point at
    which $H=0$. Thereafter, the universe ceases to expand and begins to contract.
    The contracting trajectory ends at $S$. The sudden singularity as $S$ marks both
    the beginning and end of evolution in this scenario. (The possibility that
    quantum effects might modify cosmological evolution in the vicinity of such a
    singularity has been discussed in \cite{quantum_quiescent}; see also
    \cite{nojiri1}.)

\item The universe {\em contracts\/} from the singularity at $S$ until it reaches
    $E$, where it bounces, then expands back to $S$. In this case, the brane tension
    must be negative $(\sigma < 0)$ since that is a necessary condition for moving
    along the trajectory $SE$ during contraction. In the vicinity of $\rho =
    |\sigma|$, the map $(\rho+\sigma)^2 \to \rho$ is bivalued (see
    Fig.~\ref{fig-map}), which allows $(\rho+\sigma)^2$ to increase both when $\rho$
    increases as well as decreases. This ambiguity is responsible for the two
    possibilities discussed above.

\item The trajectory $S \to P$ describes a super-accelerating universe expanding from
    the singularity $S$, since it suggests that $H^2$ increases while $\rho$
    decreases. (In fact, $\dot H \to \infty$ at the point $S$.)  If $\mu = 0$, then
    ${\dot H} > 0$ throughout this phase, and it is unlikely that $SP$ in this case
    can describe the real universe.  If $\mu > 0$, then, in the course of the
    evolution, super-acceleration may be replaced by the ``usual'' acceleration. If
    the brane tension is negative, then the point $P$ is reached, after which the
    evolution turns back to the $PS$ path.  Then, depending on the value of the brane
    tension, it either reaches the singularity $S$ again or asymptotically approaches
    the de~Sitter state at an intermediate point between $S$ and $P$.

\item For a brane with negative tension, one also has the time-reversal of the
    previous case, which describes a universe contracting either from the de~Sitter
    state at an intermediate point between $S$ and $P$ or from the singularity $S$,
    proceeding to $P$ and then to singularity $S$.

\end{enumerate}

We remember that during the evolution the parameters $A$ and $B$ describing the cubic
curve change their values, and it may happen that the curve changes its shape during the
evolution, that some critical points leave the physical domain $H^2 \ge 0$, $\rho_\t^2
\ge 0$ or, on the contrary, enter this domain.  All such possibilities are quite easy to
investigate case by case, but we will not do this in this paper.

The complete set of figures showing the $(H^2,\, \rho_{\rm tot}^2)$ plane are
shown in Fig.~\ref{fig:timelike-all} of the Appendix.

\subsection{Bouncing Braneworld}

\begin{figure}[htbp]
\begin{center}
\subfigure[]{\includegraphics[width=0.45\linewidth]{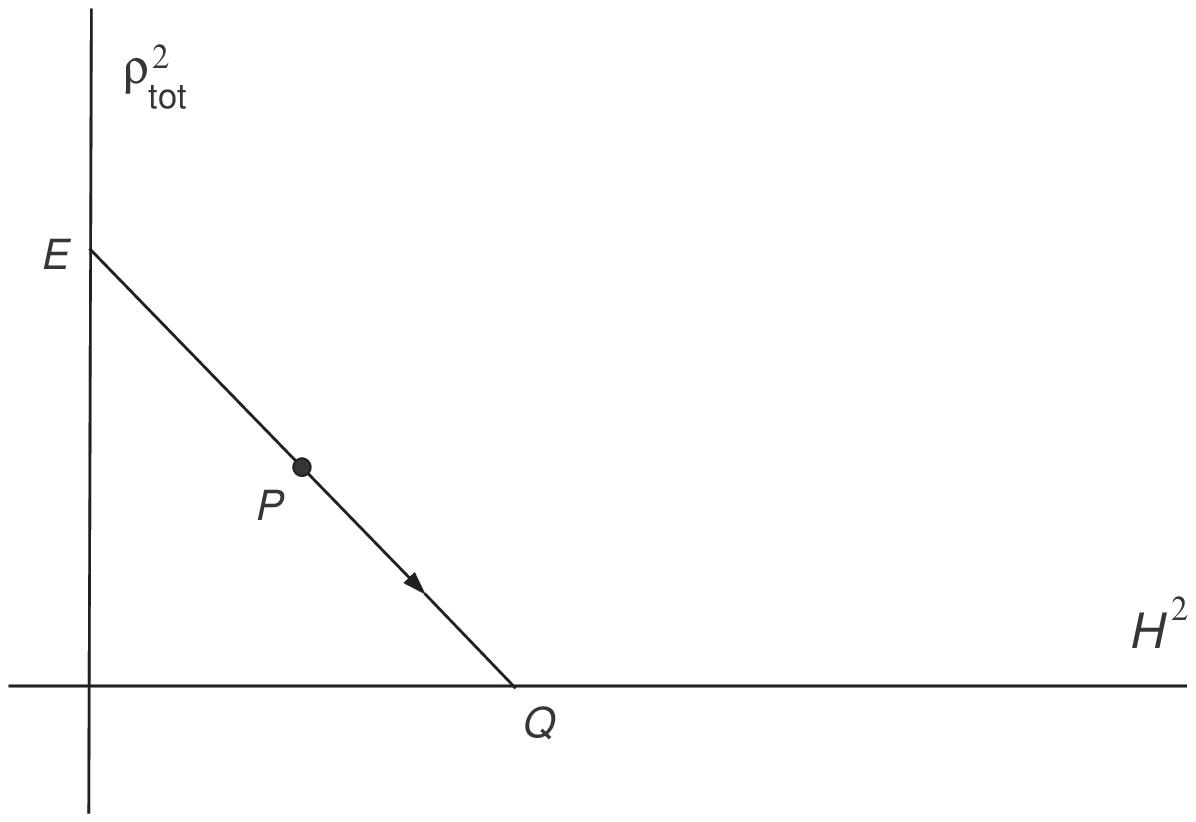}}~~~~~~~~~~~~~~~
\subfigure[]{\includegraphics[width=0.45\linewidth]{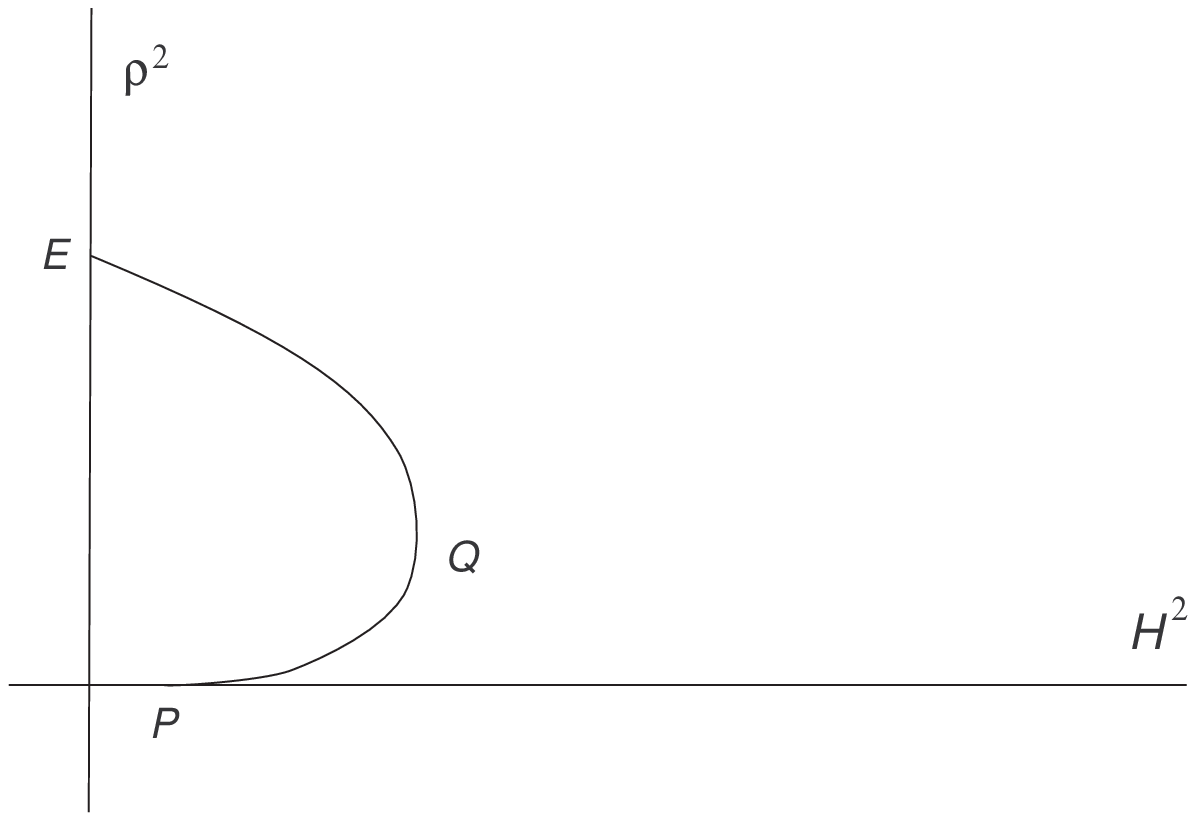}}
\caption{\label{fig-time3} The RS brane with a timelike extra dimension. Note
that the left panel shows $\rho_{\rm tot} \equiv (\rho+\sigma)^2$ as a function
of $H^2$ while the right panel shows $\rho^2$ as a function of $H^2$. Since the
map $(\rho+\sigma)^2 \to \rho$ is bivalued (see Fig.~\ref{fig-map}), it is
easier to discern the salient features of the bounce in the right panel than in
the left\,!}
\end{center}
\end{figure}

In order to address the issue of a bounce in the Gauss--Bonnet brane in more detail, let
us first consider this issue within the context of the Randall--Sundrum model (which
presents a limiting case of our braneworld). As mentioned earlier, cosmological evolution
of the RS brane is described by equation (\ref{eq:RS_bounce}) which represents a straight
line in the $\left(H^2,\, \rho_\t^2\right)$ plane. We show this line in the left panel of
Fig.~\ref{fig-time3} for a time-like extra dimension ($\varepsilon=-1$).

One can see a close qualitative similarity between the curve on the left panel of this
figure and the curve in Fig.~\ref{fig-time1}(b). The bouncing scenario in figures
\ref{fig-time1}(b) and \ref{fig-time3} proceeds as follows: the universe begins to
contract from the point $P$ at which $\rho = 0$ and $H = \mbox{const}$. In other words,
both the starting point and end point of evolution correspond to de~Sitter space. An
increase in the value of the matter density brings us to the point $Q$ at which $\rho =
|\sigma|$ and $\rho_{\rm tot}:=\rho+\sigma = 0$. (Note that $\sigma < 0$ is a
prerequisite of this model, since otherwise the point $Q$ cannot be reached.) The
universe contracts further from $Q$ to $E$ and in this segment both $\rho$ as well as
$\rho+\sigma$ increase. At $E$ the density has reached a finite maximum value while the
Hubble parameter has declined to zero. The universe therefore bounces at $E$, then
re-expands and evolves in reverse fashion along $E\to Q\to P$.

\section{Discussion}

Braneworld cosmology has attracted considerable interest during the past
decade. This is partly due to the fact that such models may play an important
role in the low energy limit of M-theory/string theory. Another reason for the
growing interest in brane dynamics is associated with the new features which
some of these models possess and which, in turn, can lead to new cosmological
predictions and scenario's. Our attempt in this paper has been to develop a
completely general qualitative approach to determine the salient features of a
brane embedded in a five dimensional bulk and evolving according to the
precepts of Einstein--Gauss--Bonnet gravity. For this purpose  we show that the
3+1 dimensional  equations of motion of several popular cosmological models can
be depicted as simple curves in the $\left(H^2,\,  (\rho+\sigma)^2\right)$
plane. (Here $H$ is the Hubble parameter, $\rho$ the density and $\sigma$ the
brane tension.)

For instance, the spatially flat FRW universe in GR has the form of a {\em
quadratic curve} while the Randall-Sundrum model describes a {\em straight
line} in the $\left(H^2,\,  (\rho+\sigma)^2\right)$ plane. The Gauss--Bonnet
brane, on the other hand, describes a {\em cubic curve} in the $\left(H^2,\,
(\rho+\sigma)^2\right)$ plane --- see equations (\ref{eq:RS_bounce}),
(\ref{eq:quadratic}) and (\ref{cubic_curve}). This {\em pictorial depiction of
dynamics\/} permits us to discover the salient features of cosmic evolution
very simply. Applying this approach to the Gauss--Bonnet brane we discover the
following interesting properties:

\begin{enumerate}

\item For a finite region in parameter space the Gauss--Bonnet brane {\em
    accelerates\/} at late times. Acceleration can be {\em phantom-like\/} ($w < -1$)
    but does not lead to the eventual destruction of the universe in a {\em
    big-rip\/} future singularity. Instead, at very late times the expansion of the
    universe approaches de~Sitter space and becomes exponential (\ie~$w \to -1$).
    (The possibility that the current expansion of the universe may be phantom-like
    has evoked much recent interest and discussion; see \cite{ss03a,phantom} for a
    non-exhaustive list of papers discussing this issue and \cite{ss06} for a summary
    of recent observational results.)

\item The expansion of the universe may commence from or terminate in a `sudden'
    quiescent singularity, at which the Hubble parameter and the density of matter
    remain finite but ${\dot H}$ diverges.

\item The universe can evade the initial big bang singularity and {\em bounce\/}.
    (This possibility is realized if the fifth dimension is timelike.)

\end{enumerate}

Whether any of these properties of the Gauss--Bonnet braneworld is realised in
practice is currently an open question which can be answered by: (i)~a deeper
understanding of the embedding of this cosmology within a more fundamental
theoretical framework, (ii)~a comparison with observations.

\section*{Acknowledgements}
HM would like to thank IUCAA, where this work was conceived and formulated, for
warm hospitality. The work of HM was supported by the Grant No.~1071125 from
FONDECYT (Chile) and the Grant-in-Aid for Scientific Research Fund of the
Ministry of Education, Culture, Sports, Science and Technology, Japan (Young
Scientists (B) 18740162). CECS is funded in part by an institutional grant from
Millennium Science Initiative, Chile, and the generous support to CECS from
Empresas CMPC is gratefully acknowledged. VS and YuS acknowledge the support of
the Indo-Ukrainian program of cooperation in science and technology sponsored
by the Department of Science and Technology of India and Ministry of Education
and Science of Ukraine.  YuS is also supported in part by the INTAS grant
No.~05-1000008-7865 and by grant No.~5-20 of the ``Cosmomicrophysics''
programme of the Ukrainian Academy of Sciences.

\vfill\eject

\section{Appendix}
\label{app:qualit}

\begin{figure}[htbp]
\begin{center}
\includegraphics[width=0.85\linewidth]{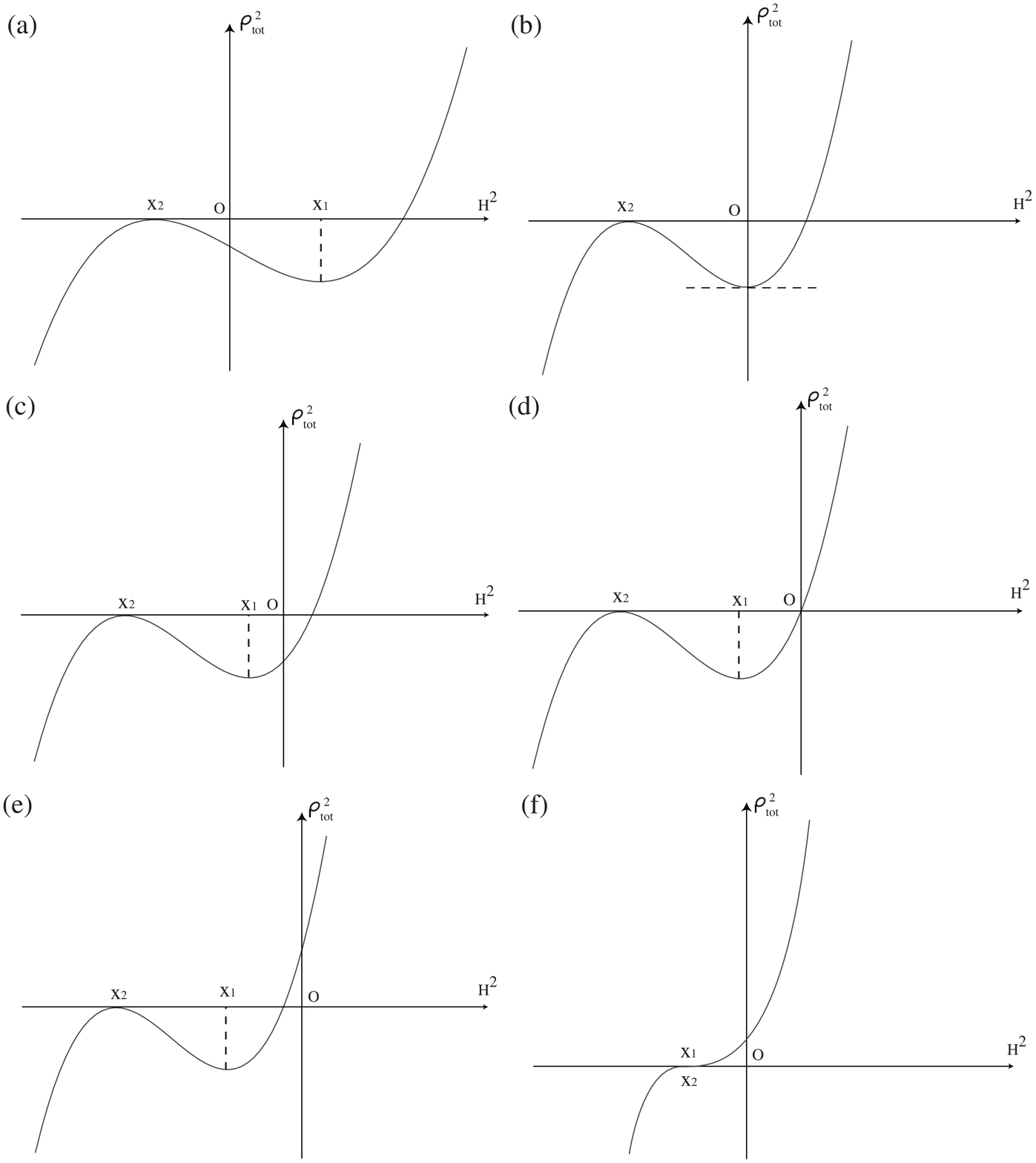}
\caption{\label{fig:spacelike-all} The $\left( H^2,\, \rho_{\rm tot}^2 \right)$
plane is shown for the GB brane with a spacelike extra dimension for the
following values of the parameters $A$, $B$ in (\ref{ab-plus}) with the upper
sign: (a)~$A < - B/2$, (b)~$A = - B/2$, (c)~$-B/2<A<0$, (d)~$A=0$, (e)~$0<A \ne
B$, and (f)~$A = B$ (at the branch singularity). Here, $x_1 := - (B + 2A)/3$
and $x_2 := -B$. Note that the region with $H^2 < 0$ or $\rho_\t^2<0$ is
nonphysical. }
\end{center}
\end{figure}

\begin{figure}[htbp]
\begin{center}
\includegraphics[width=0.37\linewidth]{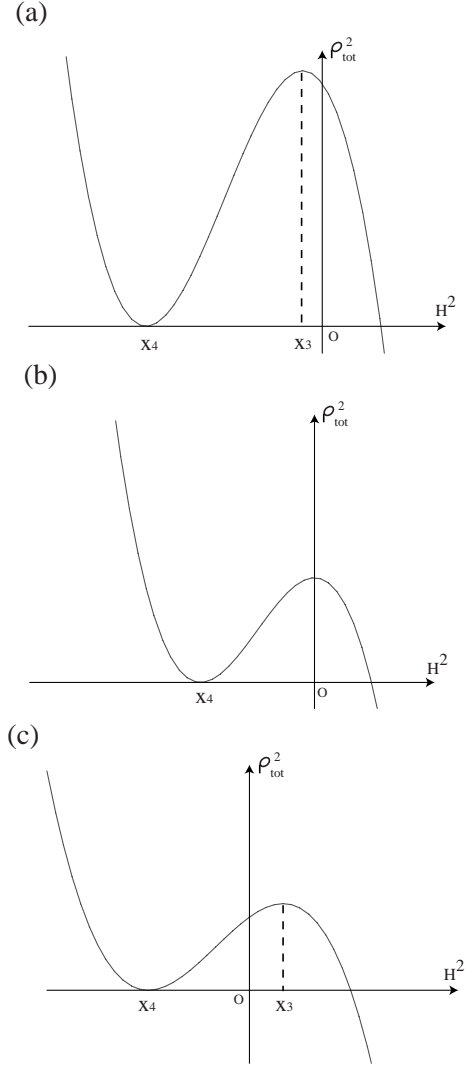}
\caption{\label{fig:timelike-all} The $\left( H^2,\, \rho_{\rm tot}^2 \right)$
plane is shown for the GB brane with a timelike extra dimension for the
following values of the parameters $A$, $B$ in (\ref{ab-minus}): (a)~$B>2A$,
(b)~$B=2A$, and (c)~$0<B<2A$. The strange singularity characterized by $A=0$ and
$B=3/(8\alpha)$ appears in the extremal case of (a). Here, $x_3:=(2A-B)/3$ and $x_4:=-B$. Note that the
region with $H^2 < 0$ or $\rho_\t^2<0$ is nonphysical. }
\end{center}
\end{figure}

Figures \ref{fig:spacelike-all} and \ref{fig:timelike-all} describe the
evolution of our brane with a spacelike and a timelike extra dimension,
respectively. These figures supplement those appearing in the main body of the
paper.

\end{document}